\documentclass[aps, prl, a4paper,showpacs,twocolumn, 10pt]{revtex4-1}
\usepackage{bbm, amsmath, amssymb, amsthm, bm,textcomp, nicefrac,geometry}
\usepackage{amsthm,amssymb,amsmath,graphicx,epsfig,color,verbatim,enumerate,amsthm}
\usepackage[dvipsnames]{xcolor}
\usepackage{ upgreek }

\newcommand{\eins}{\mathbbm{1}}

\newcommand{\ket}[1]{\left|#1\right\rangle}

\newcommand{\bra}[1]{\left\langle #1\right|}

\def\E{\mathcal{E}}
\newcommand{\be}{\begin{equation}}
\newcommand{\ee}{\end{equation}}
\newcommand{\bea}{\begin{eqnarray}}
\newcommand{\eea}{\end{eqnarray}}

\def\sone{\sigma_1}
\def\stwo{\sigma_2}
\def\sthree{\sigma_3}

\def\tr{\mathrm{tr}}

\makeatletter
\newtheorem*{rep@theorem}{\rep@title}
\newcommand{\newreptheorem}[2]{%
\newenvironment{rep#1}[1]{%
 \def\rep@title{#2 \ref{##1}}%
 \begin{rep@theorem}}%
 {\end{rep@theorem}}}
\makeatother

\newreptheorem{theorem}{Theorem}

\newtheorem*{result*}{Result}

\begin{document}
\title{Dynamical decoupling leads to improved scaling in noisy quantum metrology}
\author{P.~Sekatski, M.~Skotiniotis, and W.~D\"ur}
\affiliation{Institut f\"ur Theoretische Physik, Universit\"at Innsbruck, Technikerstr. 21a, A-6020 Innsbruck,  Austria}
\date{\today}

\begin{abstract}
We consider the usage of dynamical decoupling in quantum metrology, where the joint evolution of system plus environment is
described by a Hamiltonian. We demonstrate that by ultra-fast unitary control operations acting locally only on system qubits,
essentially all kinds of noise can be eliminated. This is done in such a way that the desired evolution is reduced by at most a
constant factor, leading to Heisenberg scaling. The only exception is noise that is generated by the Hamiltonian to be estimated
itself. However, even for such parallel noise, one can achieve an improved scaling as compared to the standard quantum limit
for any local noise by means of symmetrization.
\end{abstract}
\maketitle

%%%%%%%%%%%%%%%%%%%%%%%%%INTRODUCTION%%%%%%%%%%%%%%%%%%%%%%%%%%%
\paragraph{Introduction.---}
\label{sec:intro}
Quantum mechanics offers the promise to significantly enhance the precision of estimating unknown
parameters as compared to
any classical approach~\cite{GLM04,*GLM06,*Giovanetti:11}. Such high-precision measurements are of central importance in
physics and other areas of science, and include possible applications in frequency
standards~\cite{Wineland:92, Bollinger:96, Chwalla:07}, atomic clocks~\cite{Roos:06, Valencia:04, deburgh:05}, or gravitational
wave detectors~\cite{McKenzie:02, Ligo:11}. However, this quantum advantage seems to be rather fragile and can in general
not be maintained in the presence of incoherent noise
processes~\cite{Huelga:97,Escher:11,*Escher:12, Kn11,Kolodynski:12,*Kolodynski:13, Al14, Knysh:14}.

Identifying schemes that realize this advantage in practice are of high theoretical and practical
relevance. The usage of quantum error correction was suggested in this context, which is however restricted to certain specific
noise processes~\cite{Dur:14,*Kessler:14,*Ozeri:13, Sekatski:15}. Limited control over the environment also allows for an improved scaling in
certain situations~\cite{Gefen:15, Plenio:15}, but may be difficult to realize.

Here we provide a practical scheme to maintain quantum advantage based on the usage of dynamical decoupling
techniques~\cite{Viola:98,*Viola:99, Lidar:14}, which have been shown to be applicable for storage and for the realization of quantum gates~\cite{ Viola:09,*Viola:10}. These techniques are nowadays widely used in various experimental settings~\cite{Biercuk:09,Biercuk:09a, Sagi:10, Szwer:11, du:09, Barthel:10, bluhm:11,de:10, Ryan:10}. With the help of ultra-fast control operations that act locally on the system qubits, we show
that one can effectively decouple the system from its environment and fully protect it against decoherence effects, while at the
same time maintaining its sensing capabilities.

System and environment are described by a (pure) state, and interact via a coherent evolution governed by some Hamiltonian
$H_{SE}$. In addition, the sensing system is effected by a Hamiltonian $H_S$ that includes an unknown parameter that should
be estimated. We assume that a coherent description is appropriate at all times, and that the evolution can be interrupted by
(ultra)-fast control operations. This is similar to what is done in dynamical
decoupling~\cite{Viola:98,*Viola:99,Viola:09,*Viola:10, West:10}, or digital quantum simulation~\cite{Lloyd:96,Jane:03}.
In practice, these intermediate pulses will not be infinitely fast, and only noise up to a certain frequency can be treated and
eliminated this way. However, in what follows we will assume
that the accessible time is much faster than any other timescale in the problem.

We show that for {\em any} local system-environment interaction of rank one or two one can completely eliminate the
noise process at the cost of reducing the sensing capabilities of the system by a constant factor leading to Heiseneberg scaling
in precision. Only interactions that are of rank three---and hence necessarily contain a component that is
generated by the system Hamiltonian that should be estimated---cannot be fully corrected. Still one can achieve that
only this parallel noise part remains, and, even in this case one can still maintain a quantum advantage. In particular, for any
{\em local} noise process, we show that one can achieve a super-standard quantum scaling in precision by preparing $N$ qubit
probes in the Greenberger-Horne-Zeilinger (GHZ) state and randomizing the qubits via fast,
intermediate swap gates. For correlated noise processes we provide a general lower bound on the achievable precision,
applicable even if one assumes perfect quantum control and auxiliary resources, demonstrating that one is in general restricted
to the standard quantum limit.  Even in this case, one may still improve the precision by a constant factor, however
the exact effect depends on the details of the system-environment interactions and the type of fluctuations.

%%%%%%%%%%%%%%%%%%%%%%%%%%BACKGROUND%%%%%%%%%%%%%%%%%%%%%%%%%%%%%
\paragraph{Background.---}

Quantum metrology is the science of optimally measuring and estimating an unknown parameter such as the frequency in
an atomic clock or the strength of a magnetic field. In a typical metrology scenario, a system of $N$ probes is subjected to
an evolution for a certain time $t$ with respect to a Hamiltonian $H_S= \omega \sum \sigma_z^{(k)}$, where $\omega$ is the parameter to be
estimated, and is subsequently measured. This process is repeated $\nu$ times and $\omega$ is estimated from the
measurement statistics. The Cram\'er-Rao bound then provides a limitation on the estimation precision $\delta \omega \geq \frac{1}{\sqrt{\nu \mathcal{F}_{(t,N)}}}$, which is attainable for large enough number of repetitions $\nu$. Here ${\cal F}$ is the quantum Fisher information (QFI) ~\cite{BC94}.
For classical strategies (separable probe states) the ultimate precision is given by the standard
quantum limit (SQL), $\mathcal{F}\propto N$. While using entangled states, one can achieve the so-called
Heisenberg limit, $\mathcal{F}\propto N^2$, i.e., a quadratic improvement as compared to any classical
strategy~\cite{GLM04,*GLM06,*Giovanetti:11}.

However, in practice the system is not isolated but also interacts with its environment. In general, this leads to a noisy
evolution, where it was shown that in case of incoherent noise described by a master equation, the quantum advantage is limited
to a constant factor rather than a different
scaling~\cite{Huelga:97,Escher:11,*Escher:12, Kn11,Kolodynski:12,*Kolodynski:13, Al14, Knysh:14}. The way to describe noise
processes crucially depends on the timescales of the problem. Here we assume that we have ultra-fast access to the system,
and can interject the evolution by fast control operations, much faster than the relaxation time of the environment. In this case it is appropriate to
describe the dynamics of system plus environment in a coherent way, i.e., by means of an overall Hamiltonian that governs
the unitary evolution of system plus environment. As we will show shortly, the fact that system and environment
evolve coherently grants us with additional freedom that allows to maintain Heisenberg scaling even in the presence
of uncontrolled interaction with some environment.

%--------------------------------------------------------------------GENERAL SETTING------------------------------------------------------------------------
\paragraph*{General setting.---}We now specify the exact form of the overall Hamiltonian describing the coherent evolution of
both the system and environment.  We begin by first considering the case of a single qubit. The most general evolution
of a single qubit plus environment is described by the Hamiltonian
\be
H=H_S+H_{SE}\equiv\omega\sthree\otimes\eins + \sum_{j=0}^{3} c_j \sigma_j \otimes A_j,
\label{singlequbitH}
\ee
where $H_S=\omega\sthree\otimes\eins$ describes the evolution of the system, $H_{SE}$ describes the system-environment
interaction with $A_j$ arbitrary environment operators and $c_j$ the
coupling strengths. Here and throughout the remainder of this work $\sigma_j, \, j\in\{1,2,3\}$ denote the Pauli matrices,
${\bm\sigma}$ denotes the vector of Pauli matrices and $\sigma_{\bf n}= {\bf n}\cdot {\bm \sigma}$.

In the case where we have $N$ probe systems, the exact form of the Hamiltonian governing their coherent evolution with the
environment depends on whether the $N$ probes couple to individual environments, or to a common environment. In the former
case the Hamiltonian is given by
\begin{align}
H&=\sum_{a=1}^N \Big(\underbrace{\omega \sthree^{(a)}\otimes\eins+\sum_{j=0}^{3}c_j^{(a)}\sigma_j^{(a)}\otimes A_j^{(a)}}_{H_S^{(a)}+H_{SE}^{(a)}}\Big),
\label{independentH}
\end{align}
where $A_j^{(a)}$ act on different Hilbert spaces.  If the $N$ probes couple to a common environment then the
operators $A_j^{(a)}$ are entirely unspecified allowing
for both temporal and spatial correlations. 
Before proceeding to the results let us outline the decoupling procedure.

%--------------------------------------------------------------------DECOUPLING------------------------------------------------------------------------
\paragraph*{Decoupling strategy.---}

The most general dynamical decoupling strategy consists of applying an arbitrarily
fast time sequence of unitary gates, i.e., intersecting the system evolution with fast pulses.  Without loss of generality any strategy
corresponds to a time ordered sequence of gates $\{ u_i \}_0^n$ applied at times $\{0,t_1,\ldots,t_n\}$. If this is done fast enough, one can use a first order Trotter expansion to describe the effective evolution as being
generated by the Hamiltonian
\bea
H_\text{eff}= \mathcal{E}(H)=\sum_{i=1}^n p_i U_i H U_i^\dag
\label{effectiveH}
\eea
modulo an irrelevant final unitary, where we redefine the gates as $U_1 =u_0$, $U_{i+1} = U_i u_i$ and the probabilities are obtained from the time sequence $p_i = \frac{t_i-t_{i-1}}{t_n}$.
This is similar to what is done in optimal control theory or bang-bang control~\cite{Viola:98, Viola:99}, with the requirement that
the desired evolution $H_S$ is not completely eliminated.

\paragraph{Single qubit.---}

Notice that in the single qubit case the parametrization of completely positive trace preserving (CPTP) maps in Eq.~\eqref{effectiveH} spans exactly all the unital maps, i.e. such that
$\mathcal{E}(\eins)=\eins$~\cite{Oi:01}, and an arbitrary unital map can always be constructed from the set of generators
$\{\sigma_i\}_{i=0}^3$.  As all single qubit CPTP maps corresponds to affine transformations on the Bloch
sphere, a unital CPTP map is uniquely specified by a real matrix $A=\bar{R} D R$, where $R,\,\bar{R} \in SO(3)$,
$D=\text{diag}(\eta_1,\eta_2,\eta_3)$~\footnote{The values $\eta_i$ have to satisfy several constraints which are, however, not
important in our context}, and the action of such map on any Pauli matrix is given by
$\mathcal{E}(\sigma_{\bm n})=A{\bm n} \cdot {\bm \sigma}$. The second rotation $\bar R$ corresponds to an inconsequential change of basis, so we assume $\bar R=R^T$.

Noting that a noise term can be eliminated if and only if it belongs to the kernel of $D R$, on the other hand $D\neq 0$ since some part of the system evolution has to survive. Therefore in order to identify all the noises that can be removed it is sufficient to consider rank one projectors $A=\Uppi_{\bf r}$ in a general direction ${\bf r}=(r_1,r_2,r_3)^T$. It follows that the action of the corresponding map $\mathcal{E}=\uppi_{\bf r}$ on
Pauli matrices is given by $\uppi_{\bf r}(\sigma_{\bm n})=({\bm r} \cdot {\bm n})\, \sigma_{\bf r}$.
Notice that this remains true even if one allows for auxiliary systems and intermediate unitary operations in Eq.~
\eqref{effectiveH} to act on the enlarged system. As the argument is a purely geometrical  embedding everything in a larger 
dimensional space does not change the conclusions. Finally note that such a unital map $\uppi_{\bf r}$ can be easily 
implemented by applying $U=\sigma_{\bm r}$ at regular intervals, so that the effective Hamiltonian is
$H_{\rm eff}=\uppi_{\bf r}(H)=(H + \sigma_{\bm r} H \sigma_{\bm r})/2$.

%%%%%%%%%%%%%%%%%%%%%%%%%%%SINGLE QUBIT%%%%%%%%%%%%%%%%%%%%%%%%%%%%%

Now for the single qubit we identify all the noises that can be removed by dynamical decoupling.  As the noise term  $\eins \otimes A_0$ in Eq.~\eqref{singlequbitH} only acts on the environment and does not affect
the system evolution, it cannot be altered by any control operations performed on the system alone. In what follows we will
ignore this term, but remark that it generally plays a role for the overall evolution unless all noise
components can be canceled. The action of the decoupling strategy $\uppi_{\bf r}$ on the rest of the Hamiltonian in Eq.~\eqref{singlequbitH} leads to
\be\label{rankthreeeffective}
 H_\text{eff} =\omega \,{r}_3\,  \sigma_{\bf r} \otimes \eins + \sigma_{\bf r}\otimes \sum_{j=1}^3 { r}_j c_j A_j.
\ee Consequently the noise can be effectively decoupled if and only if
$
\exists\, \alpha_1,\alpha_2 \in \mathbb{R}$ such that $ c_3 A_3 = \sum_{j=1}^2 \alpha_j c_j A_j,
$
 in which case the effective system evolution is slowed down by the factor $r_3 =(1+\alpha_1^2+\alpha_2^2)^{-1/2}$.

In the case of bounded operators $A_j$ the Hamiltonian Eq.~\eqref{singlequbitH} can be put in the \emph{standard from}
\be
H =\omega\,  \sigma_3 \otimes \eins +\sum_{j=1}^3 b_j \sigma_{{\bm n}_j} \otimes B_j,
\ee
where $\text{tr}B_j B_k =\delta_{j k}$,  $\{{\bf n_1},{\bf n_2} ,{\bf n_3} \}$ is an orthonormal frame and $b_1 \geq b_2 \geq b_3$ are the ordered Schmidt coefficients, see supplementary material. This allows a more intuitive geometrical picture of dynamical decoupling.  For any rank one or two noise, i.e. $b_3=0$, we can choose
${\bm r}={\bm n}_1 \times {\bm n}_2$, orthogonal to the plane where noise acts. The fact that the
desired evolution $\sigma_3 \otimes \eins$ is not completely canceled  requires that ${\bm r}\cdot {\bm z} \not=0$,
which is equivalent to ${\bm n}_1,{\bm n}_2 \not= {\bm z}$.  In this case, the noise is completely eliminated by
dynamical decoupling and,as above, the system evolution
$H_S = r_3 \omega\,  \sigma_{\bf r} \otimes \eins$ is slowed down by a factor $r_3 = {\bf z} \cdot ( {\bm n}_1 \times {\bm n}_2)$.

The reduction of the coupling strength leads to a constant reduction of the
achievable accuracy by $(r_3)^2$ but, as all noise is completely eliminated, we still obtain Heisenberg scaling precision.
Notice that noise that is perpendicular to the system Hamiltonian, i.e., any combination of $\sigma_{{\bm n}_x}$ and
$\sigma_{{\bm n}_y}$ noise, can be eliminated without altering the evolution, i.e. $r_3=1$. This is done by using fast 
intermediate $\sigma_{{\bm n}_z}$ pulses.

\paragraph*{Single qubit and full rank noise.---} From the above geometrical argument, it follows that one cannot eliminate rank three noise as
such noise spans the whole three-dimensional space, and we can only eliminate a two-dimensional plane.  To see this note that
the effect of the decoupling procedure on any rank-three noise model is to project both the system
Hamiltonian and noise onto direction ${\bm r}$, so one obtains an effective Hamiltonian \eqref{rankthreeeffective} for system plus environment
which is unitarily equivalent to a $\sigma_{{\bm n}_z}$ evolution and parallel noise, with $\sum_j r_j c_j A_j= \sum_j ({\bf n}_j \cdot {\bf r})\,b_j B_j$ for the standard form.  As noise generated by the same
operator as the system Hamiltonian cannot be eliminated without eliminating the system evolution as well, the best one can
hope for in this case is to reduce any rank-three noise to noise parallel to the system evolution.

The choice of direction ${\bm r}$ in Eq.~\eqref{rankthreeeffective} determines both
the effective coupling strength of the system Hamiltonian $r_3$ and the strength of
the noise. One can then optimize ${\bm r}$ to optimize the ratio
between the modified coupling strength $r_3$  and the variance of noise fluctuations after projection \footnote{Notice,
however, that while this shows that in general we can reduce the problem to parallel noise it is not clear if this reduction to
parallel noise does in fact correspond to the optimal strategy.}.  Later on we will show what this optimal ratio is for the case of local Gaussian noise.

%%%%%%%%%%%%%%%%%%%%%%%N Qubits%%%%%%%%%%%%%%%%%%%%%%%%%%%%%%%%%%%%
\paragraph{$N$ qubits.---}

We now turn to the case of $N$ two level systems.
Consider first the case where each qubit encounters an independent environment which corresponds to a local noise
process. The total Hamiltonian describing the evolution of all $N$ qubits plus environment is given by
Eq.~\eqref{independentH}, with the system evolution $H_S=\omega \sum_a \sigma_3^{(a)} \equiv \omega S_3$, and the system-environment interaction $H_{SE} = \sum_{a=1}^N \sum_{j=1}^3  c_j^{(a)} \sigma_j^{(a)} \otimes A_j^{(a)}$ . One can use the above dynamical decoupling
strategy independently on each of the systems so that the results of the previous section directly apply. For each qubit, noise of
rank one or two can be eliminated, while full rank noise can be reduced to parallel noise  $H_{SE} = \sum_{a} \tilde c^{(a)} \sigma_{\bf r}^{(a)} \otimes \tilde A^{(a)}$ with $\tilde c^{(a)} \tilde A^{(a)}= \sum_{j=1}^3  c_j^{(a)} r_j^{(a)}  A_j^{(a)}$ (and $H_S= \omega \sum_a r_3^{(a)} \sigma_{\bf r}^{(a)}$).

In addition, one can randomize the system particles by means of fast intermediate permutations, where each permutation can
be efficiently realized by $\mathcal{O}(N)$ two-qubit swap gates.  Random permutations leave $H_S$ unchanged, but project out all
asymmetric noise terms onto their symmetric contribution~\footnote{This follows from the fact that any asymmetric operator will pick up a negative sign when
permuted by the right element of the symmetric group of $N$ objects.  As we are uniformly averaging over all elements of this
group, the only operators that will survive are the symmetric ones.}. Hence, the only remaining noise term is given by
\begin{equation}
H_{SE}=\bar{c_3} S_3 \otimes \bar{A},\quad
\bar{A}=\frac{1}{ \bar{c_3}} \sum_{a=1}^N  \tilde c^{(a)} \tilde A^{(a)}, \quad
\bar{c_3}=\frac{1}{N} \sum_{a=1}^N \tilde c^{(a)}.
\label{permnoise}
\end{equation}
Notice that in general $\bar{A}$ depends on the individual coupling strengths $\tilde c^{(a)}$ unless all $\tilde A^{(a)}$ are
identical.  As we show later symmetrization of all system qubits can, in the presence of
independent couplings or fluctuating coupling strengths, help boost precision to super-classical scaling.
We remark that if the noise has no symmetric contributions then $\bar c_3=0$, and even
locally full rank noise can be eliminated by symmetrization.

We now consider the case where the $N$ qubits couple to a common environment, which may possess both temporal and
spatial correlations. In this case the environment operators $A^{(a)}$ in Eq.~\eqref{independentH} are unspecified.  Let us first
suppose that the system-environment interactions are such that each system qubit interacts individually with the
environment.  In principle, a similar strategy as illustrated in the single-qubit case can be applied, where one
eliminates all noise except the one generated by the (symmetrized) system Hamiltonian itself by appropriate fast control
operations.  By way of example consider the following \emph{local} decoupling strategy where one applies fast
local $\sigma_z^{(a)}$ on each of the qubits.  This allows to eliminate all noise terms including $\sigma_x^{(a)},\sigma_y^{(a)}$
without altering $H_S$, and together with fast random permutations reduces all noise to one generated by the system
Hamiltonian itself, see Eq. (\ref{permnoise}).  The only difference as compared to the case of independent environments treated
above is now that the operators $\tilde A^{(a)}$ may act on the same environment.
In general a more involved decoupling strategy requiring non-local operations may be needed in order to partially or fully remove
the noise. However, it is not clear if all noise operators except those parallel to $H_S$ can be completely removed in this
case as not all unital maps can be expressed as convex combinations of unitary operations~\cite{Landau:93}. Moreover,
whatever the dynamical decoupling procedure, the condition that $H_S$ has a non-zero overlap with the kernel of the unital map
must hold in order to be able to estimate $\omega$.

One may also consider noise where several systems are affected simultaneously. From a physical standpoint
such many-body noise processes are less important as they usually correspond to higher order processes.  Nevertheless, these
correlated noise processes can be eliminated by means of dynamical decoupling, and for any quasi-local noise process
one still recovers Heisenberg scaling in the absence of noise generated by $S_3$, see supplementary material.

\paragraph{Parallel noise.---}

Hitherto, we have seen how to eliminate all kinds of noise, except noise generated by $S_3$. The latter is indistinguishable from
the desired evolution, and can not be eliminated. However, we will now show that even such parallel noise does
not automatically imply the SQL. In fact, the scaling of the QFI depends on the particular situation considered. For
instance, if the noise is due to uncorrelated fluctuations of single-qubit noise terms, then a super-SQL scaling $O(N^{3/2})$ of the QFI can
be achieved.

Consider the effect of the  system plus environment evolution described by Eq.~\eqref{permnoise} on the
system alone. Tracing out the environment in the eigenbasis $\{\ket{\ell}\}$ of $\bar A_3$ one can always represent
the noise by the CPTP map
\bea
\E(\rho)=\int p(\bar c_3) f(\ell) e^{- i t \bar c_3 \ell S_3} \rho e^{i t \bar c_3 \ell S_3} d\ell \,d \bar c_3,
\label{cptpnoise}
\eea
where $p(\bar c_3)$ corresponds to fluctuations of the interaction strength between experimental runs, and
$f(\ell) =\bra{\ell} \rho_E \ket{\ell}$ depends on the initial state of the environment~\footnote{If $\bar A_3$ has a discreet
spectrum then the integral over $\ell$ is replaced by a sum.}.

The effect of the system-environment coupling, when the environment is not in an eigenstate of $\bar A_3$, is similar to a
fluctuating interaction strength. In both cases, one has to average over evolutions governed by the same Hamiltonian as $H_S$
with a fluctuating parameter, where the latter is described by a suitable probability distribution. These fluctuations are what
ultimately limit the achievable accuracy in parameter estimation, as they directly correspond to fluctuations of the
parameter $\omega$ to be estimated. However, the resulting scaling strongly depends on the details of the situation, such as the
spectrum of environment and whether these fluctuations are correlated or uncorrelated. We now consider some of these
different cases.

The worst case is when the interaction strength, $\bar c_3$, is fixed but unknown (within a certain range). This type of noise
leads to a systematic error on the estimated value of $\omega$ and there is no way to
decrease the error below a certain value set by the initial knowledge of the interaction strength and the state of the
environment (except the trivial case where the environment is in the zero eigenstate of $\bar A_3$).

We now turn to the case where the mean interaction strength $\bar c_3$ is known but fluctuates around the mean value
between experimental runs following a smooth distribution $p(\bar c_3)$ and $\bar A_3=\eins$. This is is equivalent to the case
of a fixed $\bar c_3$ but a continuous spectrum of $\bar A_3$ with smooth $f(\ell)$.
We show in the supplementary material that for any $p(\bar c_3)$ the optimal QFI per unit time is upper bounded by
\bea\label{parallel bound}
\frac{\mathcal{F} }{t}\leq N \sqrt{\mathcal{F}_{cl}(p(\bar c_3))},
\eea
where $\mathcal{F}_{cl}(p(\bar c_3))=\int \frac{\big(p'(\bar c_3)\big)^2}{p(\bar c_3)} d\bar c_3$
remains finite for every smooth noise distribution $p(\bar c_3)$ enforcing the SQL in this case.
If $p(\bar c_3)$ is normally distributed with width $\sigma$ the bound takes the
simple form $\mathcal{F}/t \leq N/\sigma$, whereas a strategy utilizing an $N$ qubit GHZ state
$\frac{1}{\sqrt{2}}(\ket{0}^{\otimes N}+\ket{1}^{\otimes N})$  gives a maximal QFI per unit time
$\mathcal{F}/t \approx 0.43 \, N/\sigma$ for the optimal choice of $t$.

Next consider local parallel noise, where each $c_3^{(a)}$ in Eq.~\eqref{permnoise} is an independent and normally distributed
random variable with width $\sigma$.  After randomly permuting the probes
one finds that $\bar{c}_3$ is also a normally distributed random variable whose width $\bar \sigma$ is reduced by a
factor $\sqrt{N}$, $\bar{\sigma}=\frac{\sigma}{\sqrt{N}}$.  Consequently, preparing the probes in the GHZ  state yields a
super-SQL precision in estimating $\omega$
\bea
\frac{\mathcal{F}_{\mathrm{GHZ}}}{t_\text{opt}} = \frac{N^{3/2}}{\sqrt{2e}\sigma},
\eea
where $t_\text{opt}=1/\sqrt{N \sigma^2}$. Consequently, the Cram\'er-Rao bound $\delta \omega \geq (\nu \mathcal{F})^{-1/2}= (T \mathcal{F}/t_\text{opt})^{-1/2}$ is attainable for large total running time $T=\nu t_\text{opt}$.
This demonstrates that the use of symmetrization of the noise operators allows one to significantly reduce the overall effects
of noise (a fact that was also noted in~\cite{Macchiavello:02, Zanardi:99}), and restore super-SQL scaling.

Finally, in the case where $\bar c_3$ is fixed and $\bar A_3$ has a discrete spectrum, the effective noise distribution $f(\ell)$ is
discrete and $\mathcal{F}_{cl}(p(\bar c_3))$ is unbounded. Consequently, the bound of Eq.~\eqref{parallel bound} is trivial and
no general statements can be made with regards to the optimal QFI per unit time. For example, if $\bar A_3$ has an equally
gapped spectrum with gap $\Delta$, then at time $t=\frac{2\pi}{\Delta \bar c_3}$ the noise completely cancels.  This final
example, though artificial, demonstrates that one cannot provide general statements on achievable scaling without
specifying further details of the type of fluctuations, interaction, spectrum, and initial state of the environment.

\paragraph{Summary.---}

We have shown that when an overall Hamiltonian description of the system plus environment is appropriate, ultra fast control allow one to alleviate a large class of noise processes, and recover Heisenberg scaling. We remark that the dynamical procedure outlined here can also be experimentally realized with finite duration control pulses as was shown in~\cite{Viola:03}. Ultimately, the only noise
processes that forbid Heisenberg scaling precision are those generated by the system Hamiltonian to be estimated itself. The
effect of such parallel noise strongly depends on the details of interactions, the spectrum of the environment and the type of
fluctuation of the coupling parameter.

Our results are in stark contrast to situations where a master equation description of the system environment interaction is
required. There, it has been shown that with the help of auxiliary systems and fast error correction only rank one Pauli noise
processes can be eliminated, while even full quantum control including ultrafast pulses and quantum error correction do not
allow one to go beyond SQL scaling~\cite{Sekatski:15}. Hence, our results provide a big promise for practical applications of
quantum metrology in various contexts, opening the way towards ultra-sensitive devices with widespread potential application in
all branches of science.

\paragraph{Acknowledgments} We thank J. Ko\l odynski for useful discussions.
This work was supported by the Austrian Science Fund (FWF): P24273-N16, P28000-N27
and the Swiss National Science Foundation grant P2GEP2\_151964.

%----------------------------------------
\appendix
\section{Appendix}
\subsection{Standard form of the Hamiltonian}

Consider a Hamiltonian of the form
\begin{align}
H=\sum_{j=1}^{3} c_j \tilde \sigma_j \otimes A_j.
\label{independent}
\end{align}
For bounded operators $\tilde C_j = c_j A_j$ we define the overlap matrix
\be
\tilde O_{i k} = \tr \tilde C_i \tilde C_k,
\ee
which is real and symmetric $\tilde O=\tilde O^T$ (as imposed by the hermiticity of the Hamiltonian $\tilde C_j=\tilde C_j^\dag$).
Expressing the Pauli operators in a rotated frame $\tilde {\bm \sigma}= R\,  {\bm \sigma}$ allows one to rewrite the Hamiltonian as
\be
H=\sum_{k=1}^{3} \sigma_k \otimes \sum_j R_{j k}\tilde C_j = \sum_{k=1}^{3} \sigma_k \otimes C_k,
\ee
with $C_k = \sum_{j=1}^3 R_{jk}\tilde C_{k}$. Accordingly the overlap matrix for the operators $O_{i k} = \tr  C_i C_k$ is given by
\be
O = R^T \tilde O R.
\ee
Choosing the rotation that diagonalizes the symmetric matrix $\tilde O= R \,\text{diag}(\lambda_1,\lambda_2,\lambda_3) \, R^T$ leads to
\be
\tr C_j C_k = \delta_{jk} \lambda_j.
\ee
Which also shows that $\lambda_j \geq 0$, being the trace of the square of an Hermitian operator. Finally, denoting
 $B_j = \frac{1}{\sqrt{\lambda_j}} C_j$ and $b_j= \sqrt{\lambda_j}$ allows one to rewrite the Hamiltonian as
\be
H= \sum_{j=1}^3 b_j \,\sigma_j \otimes B_j,
\ee
with $\tr B_j B_k =\delta_{jk}$.

\subsection{Correlated noise}
\label{correlated}
We now consider correlated noise processes where several systems are affected jointly. In general, the
system-environment Hamiltonian of a $N$-qubit system is given by
$H_{SE} = \sum_{{\bm j}} c_{\bm j}^{({\bm a})} T_{\bm j}^{({\bm a})} \otimes A_{\bm j}^{({\bm a})}$ where
$T_{\bm j}^{({\bm a})} =\sigma_{j_1}^{(a_1)}\otimes \sigma_{j_2}^{(a_2)} \ldots \sigma_{j_N}^{(a_N)}$ denotes a tensor product
of Pauli operators. Using fast intermediate $\sthree$ operations on all qubits allows one to eliminate all terms containing
$\sone,\,\stwo$ somewhere. We are then left with a Hamiltonian where $j_k \in \{0,3\}$ and noise is solely diagonal.
In case of localized noise, i.e., where there is a certain spatial structure and only qubits that are spatially close are jointly
affected by noise, one can use fast intermediate $\sone$ operations acting sparsely to eliminate noise terms of range $k$.
For instance, performing such an action on every second qubit eliminates all nearest neighbor two-qubit noise terms in a
$1$-D setting. However, this also eliminates the desired evolution for half of the particles, and these particles no longer
contribute to the sensing process. As long as the number of systems to be decoupled is given by $\alpha N$ with $\alpha$
being some constant---which is the case for any finite range $k$ noise operators---we still obtain Heisenberg scaling
${\cal O}(\alpha^2 N^2)$.

\subsection{{Parallel noise upper-bound QFI}}
\label{Parallel}
In this section we derive a limitation on the maximally achievable QFI in presence of the parallel noise, i.e., noise that is described by the same generator as  the Hamiltonian $H$ that governs the evolution. Such parallel noise results in the
channel
\bea
\label{parallel noise}
\E(\rho) = \int p(\lambda) e^{-i t \lambda H } \rho \,e^{i t \lambda H } d\lambda,
\eea
where $\lambda$ is a random variable and $p(\lambda)$ is a probability distribution with standard deviation $\sigma$
characterizing the strength of the noise.  As already mentioned such type of noise cannot be ameliorated using error
correction as the operator generating it is identical to the Hamiltonian generating the desired evolution.  The noise process
in Eq.~\eqref{parallel noise} can be viewed as describing classical noise applied directly on the estimated parameter $\omega$, i.e.,
in every run of the experiment the observed parameter fluctuates by an amount $\lambda$, with $\lambda$ being a random
variable with corresponding probability distribution  $p(\lambda)$.

Recall that the QFI of a state $\rho$  is given by~\cite{BC94}
\bea
\mathcal{F}\Big(\rho(\theta)\Big) = 8 \frac{1-F\Big(\rho(\theta),\rho(\theta+d\theta)\Big)}{d\theta^2},
\label{A2}
\eea
where $\rho(\theta) = e^{-i \theta H } \rho \,e^{i \theta H }$ and $F(\rho,\tau) = \tr \sqrt{\tau^{1/2} \rho\, \tau^{1/2}}$ is the Uhlmann fidelity. So one can access the QFI in the presence of parallel noise (Eq.~\eqref{parallel noise}) through the Uhlmann fidelity
\begin{align}
F\Big(\E(\rho(t \omega)),\E(\rho(t \omega+ t d\omega))\Big)= \nonumber\\ F\Big(  \int p(\lambda) \rho(t \omega +t \lambda)  d\lambda,
 \int p(\lambda-d\omega) \rho(t \omega+ t \lambda) d\lambda\Big).
 \label{A3}
\end{align}
As the Uhlmann fidelity is strongly concave it follows that Eq.~\eqref{A3} is lower bounded by the fidelity of the probability
distributions $p(\lambda)$ and $p(\lambda +d\omega)$. Consequently the QFI in the presence of parallel noise is bounded by
\bea
 \mathcal{F}\Big(\mathcal{E}(\rho)\Big)\leq \int \frac{(p'(\lambda))^2}{p(\lambda)}d\lambda =\mathcal{F}_{cl}\Big(p(\lambda)
 \Big).
\eea

On the other hand we know that the QFI in the noisy case is lower that the noiseless QFI  (atteined by the GHZ state), therefore $\mathcal{F}\Big(\E(\ket{\xi})\Big)\leq t^2 N^2$. Combining the two bounds one gets for the QFI per unit time
\bea
\frac{\mathcal{F}\Big(\E(\ket{\xi})\Big)}{t} \leq \min\left(t N^2, \frac{ \mathcal{F}_{cl}\Big(p(\lambda)\Big)}{t}\right).
\eea
It remains to find the time $t$ that maximizes the r.h.s. Trivially the maximum is attained when $t N^2 =  \frac{ \mathcal{F}_{cl}(p(\lambda))}{t}$, which yields
\bea
\frac{\mathcal{F}\Big(\E(\ket{\xi})\Big)}{t} \leq  N \sqrt{ \mathcal{F}_{cl}\Big(p(\lambda)\Big)}.
\eea

For any smooth distribution $p(\lambda)$ the classical Fisher information $ \mathcal{F}_{cl}(p(\lambda))$ is finite, and therefore SQL scaling for the QFI per unit time is enforced. In particular for a Gaussian noise with $p(\lambda)= \frac{1}{\sqrt{2\pi\sigma^2}} e^{- \lambda^2/2\sigma^2}$ this bound implies
\bea
\frac{\mathcal{F}\Big(\E(\ket{\xi})\Big)}{t} \leq \frac{N}{\sigma}.
\eea
While for a simple strategy with GHZ states and the optimal choice of the time a straightforward calculation gives $\mathcal{F}/t \approx 0.429 N/\sigma$, which is roughly half of the bound above.

\bibliographystyle{apsrev4-1}
\bibliography{ultimatelimits}

\end{document}